\documentclass[mathleft,fleqn,%
]{an}
\usepackage{graphicx}
\usepackage[varg]{txfonts}
\begin{document}

\Pagespan{1}{}
\Yearpublication{2014}%
\Yearsubmission{2014}%
\Month{0}%
\Volume{999}%
\Issue{0}%
\DOI{asna.201400000}%

\title{Determining rotational and macroturbulent velocities of cool magnetic giant stars\thanks{Based on data obtained using the T{\'e}lescope Bernard Lyot at Observatoire du Pic du Midi, CNRS and Universit{\'e} de Toulouse, France.}}

\author{S.\,~Georgiev\inst{1,2}\fnmsep\thanks{Corresponding author:
        {sgeorgiev@astro.bas.bg}}
A.\,~L{\`e}bre\inst{2}, E.\,~Josselin\inst{2}, R.\,~Konstantinova-Antova\inst{1}, J.\,~Morin\inst{2}
}
\titlerunning{Determining rotational and macroturbulent velocities of cool magnetic giant stars}
\authorrunning{S.\, Georgiev et al.}
\institute{
Institute of Astronomy and NAO, Bulgarian Academy of Science, 72, Tsarigradsko Chaussee Blvd., 1784 Sofia, Bulgaria
\and 
LUPM, UMR 5299, Universit{\'e} de Montpellier, CNRS, place Eug{\`e}ne Bataillon, 34095 Montpellier, France}

\received{XXXX}
\accepted{XXXX}
\publonline{XXXX}


\abstract{%
An original method of estimating the projected rotational velocity, $vsini$, and the macroturbulent velocity, $v_{\rm mac}$, of evolved M giant stars is presented. It is based on the use of spectrum synthesis and multi-line analysis tools. The goal is to fit the mean line profile of observations with that of synthetic spectra. The method is applied to the red giant star RZ~Ari and the results $v\sin i$ = 6.0 $\pm$ 0.5 km/s and $v_{\rm mac}$ = 2.0 $\pm$ 1.0 km/s are obtained.}

\maketitle

\section{Introduction}
RZ~Ari (HD~18191) is a magnetically active M6\,III semi-regular variable star (see Konstantinova-Antova et al. 2018 and references therein). The physical scenario behind the magnetic variability of this star is still unclear, and so is the role of rotation in it. The rotational period of RZ Ari is not precisely known, as is the case for most M~giants. Based on results of Zeeman-Doppler Imaging (ZDI, Semel 1989; Donati \& Brown 1997; Donati et al. 2006) and period analysis of spectral activity indicators, Konstantinova-Antova et al. (2020, in prep.) suggest that for this star the rotational period should be between 700 and 1100 d. On the other hand, the only estimations of the projected rotational velocity of RZ Ari available in the literature are given by Zamanov et al. (2008), who obtain $v\sin i = 9.6 \pm 2.0$~km/s by means of cross-correlation and $v\sin i = 12.0 \pm 2.0$~km/s by measuring the full width at half-maximum (FWHM) of spectral lines (both methods of estimation are described in Zamanov et al. 2007). However, we find that these values are not compatible with observations when we attempt to model the unpolarized and circularly polarized profiles of spectral lines of RZ~Ari using the ZDI method. We suspect that these estimations of $v\sin i$ are in fact upper limits of the real value, because they were done without properly considering the effect of macroturbulence: as Gray (2013) shows, the use of cross-correlation techniques to estimate $v\sin i$ of cool stars bears uncertainties due to the handling of line broadening caused by macroturbulence. In order to constrain the origin of the detected magnetic field at the surface of RZ~Ari, and its connection with rotation, a better estimation of the rotational velocity that properly takes into account the effects of macroturbulence is needed. Such an estimation is the goal of the present work.\\
A classical approach for the determination of stellar parameters, including $v\sin i$ and the macroturbulent velocity, $v_{\rm mac}$ is the spectrum synthesis method, which relies on high-resolution spectroscopic observations. Stellar model atmospheres and atomic and molecular linelists are required to create a synthetic spectrum. The latter is convoluted with the instrumental profile of the spectrograph used for the observations and also with profiles of rotational and macroturbulent broadening. It is then compared, over a large spectral domain, to an observed spectrum to obtain the fundamental parameters (including $v\sin i$ and $v_{\rm mac}$) of the considered star. However, in the case of cool M giant stars one faces certain difficulties in applying this method, such as the great number of spectral features that often blend each other and the likely imperfect normalization to the continuum often affecting the observation and preventing a straightforward comparison with a synthetic spectrum.\\
Another method for estimating stellar parameters, including $v\sin i$, is using the cross-correlation technique. For example, the instrument SOPHIE (Perruchot et al. 2008; Bouchy et al. 2009) can obtain high-resolution stellar spectra and calculate the cross-correlation function (CCF) of the observed spectrum and a numerical mask. The CCF is then fit with a gaussian profile and stellar parameters are derived from its parameters. A detailed explanation of the cross-correlation technique can be found in Melo et al. (2001).\\
To estimate the rotational and macroturbulent velocities in the case of the cool M~giant RZ~Ari, we propose to use an original multi-line approach, using the Least Square Deconvolution method (LSD, Donati et al. 1997). LSD works under the assumption that all spectral lines excluding very strong ones (like the Balmer or CaII H{\&}K lines) are similar in shape and simply scale up in depth. Using LSD, a mean spectral line profile is produced for both observational and synthetic data. By doing this, we can base our comparison on an observational and a synthetic mean spectral line that encode the fundamental stellar parameters for several different pairs of $v\sin i$ and $v_{\rm mac}$. Because thousands of spectral lines are considered in the calculation of the mean profile, the comparison is performed with observational data with significantly increased signal/noise ratio (SNR) with respect to that of the spectra. Working with the mean line profiles instead of the full spectra also has the advantage of simplicity and allows some degree of automation. The latter provides possibility for working with large datasets to obtain a more statistically significant result. This method is original in the sense that LSD has not been used together with spectrum synthesis to obtain stellar parameters before. In contrast with cross-correlation, the parameters are not derived from a simple gaussian fit to the mean line profile.\\
\section{Observations and LSD software}
Observations of RZ~Ari were done using the spectropolarimeter Narval on the 2m Telescope Bernard-Lyot (TBL) at the Pic du Midi observatory, France. The Narval instrument (Auri{\`e}re 2003) has a resolving power of 65000 and operates in the spectral range of 375 to 1050 nm. It allows simultaneous measurement of the full intensity (Stokes~I) and the intensity in linear (Stokes~U or Q) or circular (Stokes~V) polarization as a function of wavelength.\\
RZ~Ari has been observed (in circular polarization) during  the period September 2008 - March 2019. The 57 collected spectra were treated initially by the automatic software LibreEsprit (Donati et al. 1997) that performs the spectrum extraction, wavelength calibration, heliocentric frame correction and continuum normalization. A log of observations is presented in Table 1.
\begin{table}[]
\resizebox{\columnwidth}{!}{\begin{tabular}{lll|lll}
\hline
DATE    & \begin{tabular}[c]{@{}l@{}}HJD\\ 2450000+\end{tabular} & \begin{tabular}[c]{@{}l@{}}Mean SNR\\ 400-550 nm\end{tabular} & DATE    & \begin{tabular}[c]{@{}l@{}}HJD\\ 2450000+\end{tabular} & \begin{tabular}[c]{@{}l@{}}Mean SNR\\ 400-550 nm\end{tabular} \\ \hline
16sep08 & 4727                                                   & 249                                                             & 03dec13 & 6630                                                   & 168                                                             \\
21sep08 & 4732                                                   & 215                                                             & 09jan14 & 6667                                                   & 241                                                             \\
05sep10 & 5446                                                   & 243                                                             & 19aug15 & 7255                                                   & 179                                                             \\
21sep10 & 5462                                                   & 167                                                             & 05sep15 & 7272                                                   & 178                                                             \\
13oct10 & 5484                                                   & 280                                                             & 08oct15 & 7305                                                   & 197                                                             \\
22jan11 & 5584                                                   & 127                                                             & 31oct15 & 7328                                                   & 179                                                             \\
27jan11 & 5589                                                   & 174                                                             & 30nov15 & 7358                                                   & 165                                                             \\
04feb11 & 5597                                                   & 224                                                             & 18dec15 & 7375                                                   & 201                                                             \\
26sep11 & 5832                                                   & 177                                                             & 05aug16 & 7607                                                   & 179                                                             \\
16oct11 & 5851                                                   & 236                                                             & 01sep16 & 7634                                                   & 208                                                             \\
23nov11 & 5889                                                   & 187                                                             & 03oct16 & 7666                                                   & 205                                                             \\
24nov11 & 5890                                                   & 177                                                             & 29oct16 & 7691                                                   & 153                                                             \\
10jan12 & 5937                                                   & 236                                                             & 01dec16 & 7724                                                   & 144                                                             \\
11jan12 & 5938                                                   & 218                                                             & 20dec16 & 7743                                                   & 109                                                             \\
16jul12 & 6126                                                   & 224                                                             & 07jan17 & 7761                                                   & 200                                                             \\
17jul12 & 6127                                                   & 214                                                             & 16feb17 & 7801                                                   & 208                                                             \\
18jul12 & 6128                                                   & 146                                                             & 04sep17 & 8002                                                   & 114                                                             \\
16aug12 & 6157                                                   & 164                                                             & 06oct17 & 8034                                                   & 184                                                             \\
17aug12 & 6158                                                   & 169                                                             & 30oct17 & 8057                                                   & 204                                                             \\
04sep12 & 6176                                                   & 225                                                             & 23nov17 & 8081                                                   & 116                                                             \\
05sep12 & 6177                                                   & 251                                                             & 23jan18 & 8142                                                   & 103                                                             \\
04oct12 & 6206                                                   & 206                                                             & 18sep18 & 8381                                                   & 154                                                             \\
12nov12 & 6244                                                   & 170                                                             & 22oct18 & 8415                                                   & 186                                                             \\
11jan13 & 6304                                                   & 202                                                             & 16nov18 & 8439                                                   & 167                                                             \\
08jul13 & 6483                                                   & 176                                                             & 07jan19 & 8491                                                   & 194                                                             \\
05aug13 & 6511                                                   & 183                                                             & 26jan19 & 8510                                                   & 108                                                             \\
02sep13 & 6539                                                   & 207                                                             & 08mar19 & 8551                                                   & 119                                                             \\
06oct13 & 6573                                                   & 230                                                             & 11mar19 & 8554                                                   & 166                                                             \\
07nov13 & 6604                                                   & 165                                                             &         &                                                        &                                                                 \\ \hline
\end{tabular}}
\caption{Log of observations of RZ~Ari. The calendar date, heliocentric julian date (HJD) and mean signal/noise ratio (SNR) in the spectral window used in the estimation of $v\sin i$ and $v_{\rm mac}$ are shown in their respective columns.}
\end{table}
\\The LSD software uses a line mask to average the profiles of a large number of atomic lines, assuming that they have the same shape scaled by a certain factor. The line mask is a reference line pattern that indicates the laboratory wavelengths of spectral lines along with the atomic number of the chemical element they are associated with, their local depth (as a percentage of local continuum level), excitation potential and Land{\'e} factor. The mask is computed by solving the radiative transfer equation for a model atmosphere using atomic and molecular linelists. LSD then performs a cross-correlation of the observed spectrum with the line mask to compute the mean line profile (the reader can refer to Section 4 of Donati et al. 1997 for an in-depth description of the LSD method). The result is a mean line profile (called an LSD profile) in heliocentric velocity space in both Stokes~I and polarized light (Stokes~V for our observations). This tool is widely used in high-resolution spectropolarimetry to detect mean polarization signatures with very high SNR. For example, in the case of RZ~Ari it was by detecting a strong signal in the LSD Stokes~V profiles that the presence of a surface magnetic field was detected (Konstantinova-Antova et al. 2013). In the present study however, only the LSD Stokes~I profile is used. To perform LSD on RZ~Ari data, we used a mask computed using a MARCS model atmosphere (Gustafsson et al. 2008) and atomic and molecular line lists extracted from the Vienna Atomic Line Database VALD (Kupka et al. 1999) for the following stellar parameters: $T_{\rm eff}$ = 3400K, $\log g$ = 0.5, microturbulent velocity of 2~km/s. The mask covers ${\approx}19000$ lines in the full spectral domain of Narval.

\section{Determination of $v\sin i$ and $v_{\rm mac}$}
Our goal is to determine the $v\sin i$ and $v_{\rm mac}$ parameters by comparing the Stokes~I LSD profiles of observations to those of synthetic spectra. To create the synthetic spectra, first a model of the stellar atmosphere is necessary.\\
Different values for the stellar parameters have been reported for RZ Ari: $T_{\rm eff} = 3442$~K (van Dyck et al. 1998), $T_{\rm eff} = 3450$~K (Konstantinova-Antova et al. 2010), $T_{\rm eff} = 3250$~K, $\log g = 0.30$ and [Fe/H] = $-0.24$ (Prugniel et al 2011). We used a MARCS model atmosphere with standard composition in spherical model geometry with the following stellar parameters: $T_{\rm eff} = 3400$~K, $\log g$ = 0.5, [Fe/H] = -0.25, $v_{\rm mic} = 2$~km/s, $M = 1M_{\odot}$. We chose to work with an effective temperature of 3400K because the difference in the normalization to the continuum between our observations and the synthetic spectra is smallest for this $T_{\rm eff}$.\\
Using this model atmosphere, a synthetic spectrum was constructed with the software Turbospectrum (Plez 2012) that solves the radiative transfer equation using atomic and molecular line lists extracted from VALD. The computations were limited in wavelength between 400 and 550 nm with a step of $10^{-3}$ nm. The choice of this spectral window is justified by three facts: first, the normalization to the continuum of the observational data performed by LibreEsprit is done best in the blue part of the spectrum for very cool stars, while in the red part the presence of many molecular bands (mainly of TiO) prevents a good normalization to the continuum; second, the typical SNR of the observations is sufficiently high (${\approx}190$) through this spectral window, allowing reliable comparison; third, the density of atomic lines in the LSD mask is higher in the blue part of the spectrum. The observations were also cut to match this spectral region.\\
Having created the synthetic spectrum, it is necessary to take into account the instrumental profile of Narval, which we consider to be a gaussian with a FWHM of 77 m$\AA$. Turbospectrum has a built-in procedure for performing this step, which consists of doing a convolution of the spectrum with the instrumental profile.\\
Next, the effects of macroturbulence and rotation have to be considered. This is achieved again using the procedure implemented in Turbospectrum which performs the convolution of the synthetic spectrum with the two corresponding profiles, in successive operations. We used a radial-tangential profile to model the macroturbulence and a rotational profile, both as described in Gray (1992). Then, a grid of synthetic spectra was produced, covering the range of $v_{\rm mac}$ between 1 and 6 km/s with a step of 1 km/s and $v\sin i$ between 4.5 and 10 km/s with a step of 0.5 km/s. Finally, a Doppler velocity correction was applied to take into account the radial velocity of RZ~Ari ($46.5 \pm 0.9$ km/s as measured from the observational data). In this way, we obtained what we call a grid of synthetic Narval observations.\\
The LSD software was then applied to this grid of synthetic data using the same line mask as for the Narval observations. The typical number of lines used by LSD when working with the 400-550 nm spectra was ${\approx}7000$.\\
The mean spectral line profile computed for each synthetic data in the grid was then compared to that of each individual observation. The LSD output gives the light intensity, $I$, as a function of heliocentric velocity, $V$. Since the LSD procedure was performed in the exact same way for all the data (synthetic and observational ones), the distribution of points in the velocity space is uniform for the profiles, beginning at -243 km/s and ending at 243 km/s with a step of 1.8 km/s. This means that comparison between the  different mean line profiles can be done simply by comparing the differences between their individual intensities $I(V)$ that share the same velocity coordinates. To estimate the correspondance between each pair of observational and synthetic LSD profiles we calculate the sum
\begin{equation}
S = {\sum}_{V = V_{min}}^{V_{max}} (I_{syn}(V) - I_{obs}(V))^2
\end{equation}
The criterium to select the best fit of the LSD profile for each observation is the minimal sum for all synthetic observations in the grid, $S = S_{min}$.\\
The values $V_{min}$ and $V_{max}$ are calculated by fitting the observed mean line profile with a gaussian function with a peak at $V_0$ and FWHM = ${\sigma}$. We then set $V_{min} = V_0 - 1.5{\sigma}$ and $V_{max} = V_0$, i.e. we only consider the blue part of the mean line profile (see Figure 1). Initially, the whole line profile was used; however, this approach did not lead to convergence of the results for the $v\sin i$ and $v_{\rm mac}$ parameters. As one can see in Figure 1, the red wing of the observational mean line profile in the 400-550 nm region displays a dip in intensity which is never matched by the synthetic mean line profile and is not seen in the blue wing. This asymmetric shape is not found when examining the observational mean line profiles of the full 375-1050 nm spectra. The blue wing however appears consistent between the mean profiles of the cut and full spectra and it is also well reproduced by the synthetic observations, which is why only it was used. The lines used in our analysis are in the blue part of the visible domain. On average,  they have higher excitation potentials; we think that if these lines (formed deeper in the atmosphere) are sensitive to downward motions, this could distort their profiles, especially in the red part. We suspect that this effect might to some extent explain the asymmetry between the blue and red wings of the mean line profiles. The asymmetry could then disappear when using the full visible domain, since the effect would be cancelled out when lower excitation lines in the red are introduced.
\begin{figure}
\resizebox{\columnwidth}{!}{{\includegraphics{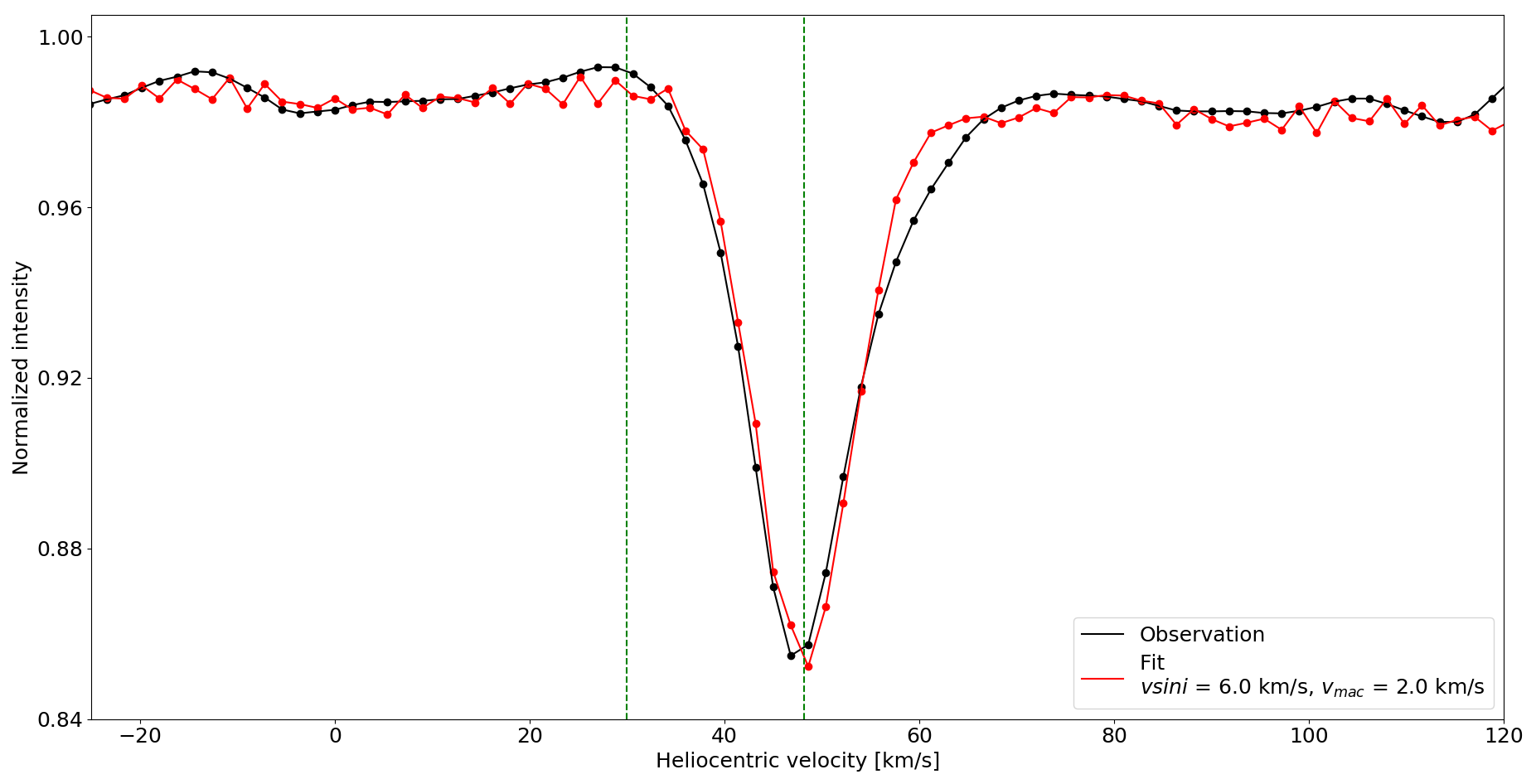}}}
\caption{Example of the comparison between LSD Stokes~I profiles of observed (black line) and synthetic spectra (red line). The observation is collected on 2016 December 20. The vertical green dashed lines show the range in which the evaluation sum has been calculated. Note the asymmetry between the blue and red wings of the observational mean line profile (see the text).}
\end{figure}
\\After obtaining the best results according to the minimization rule, the fits were examined individually. 16 (out of the 57) dates showed LSD profiles significantly different from the other mean line profiles and could not be correctly fit by the synthetic data. We think that this is due to the physical variability of RZ~Ari. These outlying dates were not considered in the final estimation of $v\sin i$ and v$_{mac}$.
\\For each of the 41 observations that were well fit by the synthetic data, the best fitting ($v\sin i$, $v_{\rm mac}$) pair was noted on a two-dimentional color plot with the $v\sin i$ and $v_{\rm mac}$ on the horizontal and vertical axis respectively. The plot is shown in Figure 2, where the color code represents the number of best fits obtained for each ($v\sin i$, $v_{\rm mac}$) pair.
\begin{figure}
\resizebox{\columnwidth}{!}{{\includegraphics{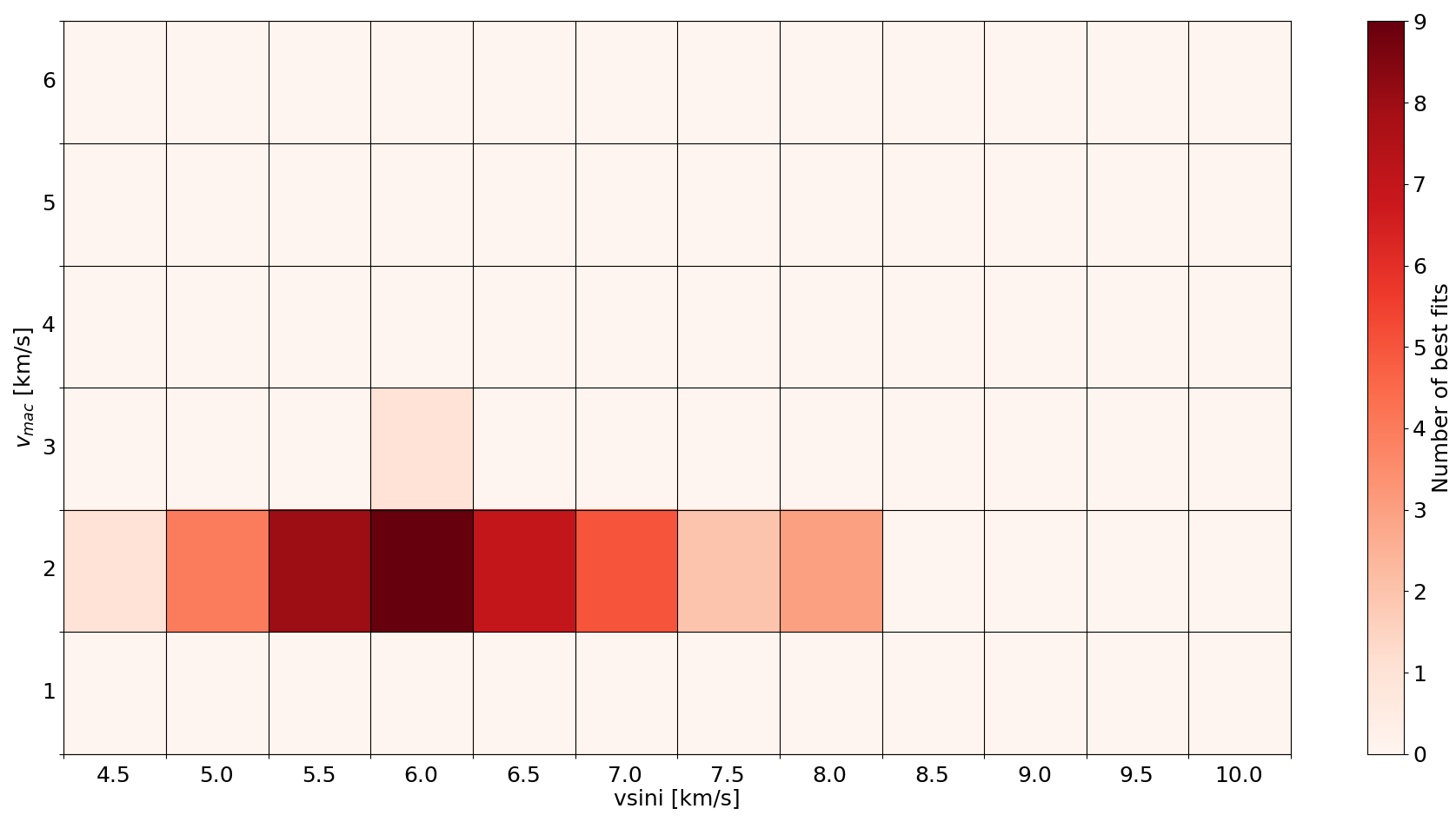}}}
\caption{Distribution of the results for the 41 well fit observations with respect to $v\sin i$ and $v_{\rm mac}$.}
\end{figure}
\\From the distribution presented in Figure 2, it is clear that the best fits are most often obtained with the set of parameters $v_{\rm mac}=2.0$ km/s and $v\sin i=6.0$ km/s (9 best fits), 5.5 km/s (8) and 6.5 km/s (7). Other, less populated peaks in the distribution are seen at $v\sin i=7.0$ km/s (5), 5.0 km/s (4) for the same macroturbulent velocity. Individual fits that match the minimization criterium appear for other combinations of the two parameters. Except a single observation that is fit best with $v_{\rm mac}=3.0$ km/s and $v\sin i=6.0$ km/s, no result is obtained with $v_{\rm mac}{\neq}2.0$~km/s.\\
Based on the statistical distribution presented in Figure 2, we estimate for RZ~Ari the values of $v_{\rm mac}=2.0{\pm}1.0$~km/s and $v\sin i=6.0{\pm}0.5$~km/s, where the errorbars correspond to the steps in the grid we used. Such an estimation of the error is obviously very rough. The uncertainties in the selected values of $T_{\rm eff}$, $\log g$, [Fe/H] and microturbulence used when computing the synthetic spectrum and line mask all impact the errorbars of $v\sin i$ and $v_{\rm mac}$. It must also be metioned that the precision of the estimated parameters depends on the step of the grid of the model atmospheres. For cool stars ($T_{\rm eff} \leq 4000$~K) the steps in the MARCS models for, e.g. $T_{\rm eff}$ and $\log g$ are respectively $100$K and $0.5$. In a subsequent study, a more quantitative error estimation will  be required.\\\\
To test the method, we also used it to estimate the $v\sin i$ and $v_{\rm mac}$ parameters of another star, EK~Boo (HD~130144), which is of spectral type M5\,III, very similar to that of RZ~Ari. Konstantinova-Antova et al. (2010) report for EK~Boo the values $v\sin i=8.5\pm0.5$~km/s and $v_{\rm mac}=2.0$~km/s obtained using classical spectrum synthesis. For our calculations we used a mask computed using line lists extracted from VALD for the following parameters: $T_{\rm eff}=3500$~K, $\log g=0.5$, microturbulent velocity of 2~km/s. The MARCS model atmosphere used to create the synthetic spectrum was computed using standard composition in spherical model geometry with $T_{\rm eff}=3500$~K, $\log g=0.5$, [Fe/H] = 0, microturbulent velocity of 2~km/s, $M=1M_{\odot}$. We used 54 Narval observations of EK~Boo restricted to the same spectral window (400-550 nm). Following the same minimization criterium as in the case of RZ~Ari, we obtain the results shown in Figure 3 and conclude the values $v\sin i=8.0{\pm}0.5$~km/s, $v_{\rm mac}=1.0{\pm}1.0$~km/s, in good agreement with Konstantinova-Antova et al. (2010). 
\begin{figure}
\resizebox{\columnwidth}{!}{{\includegraphics{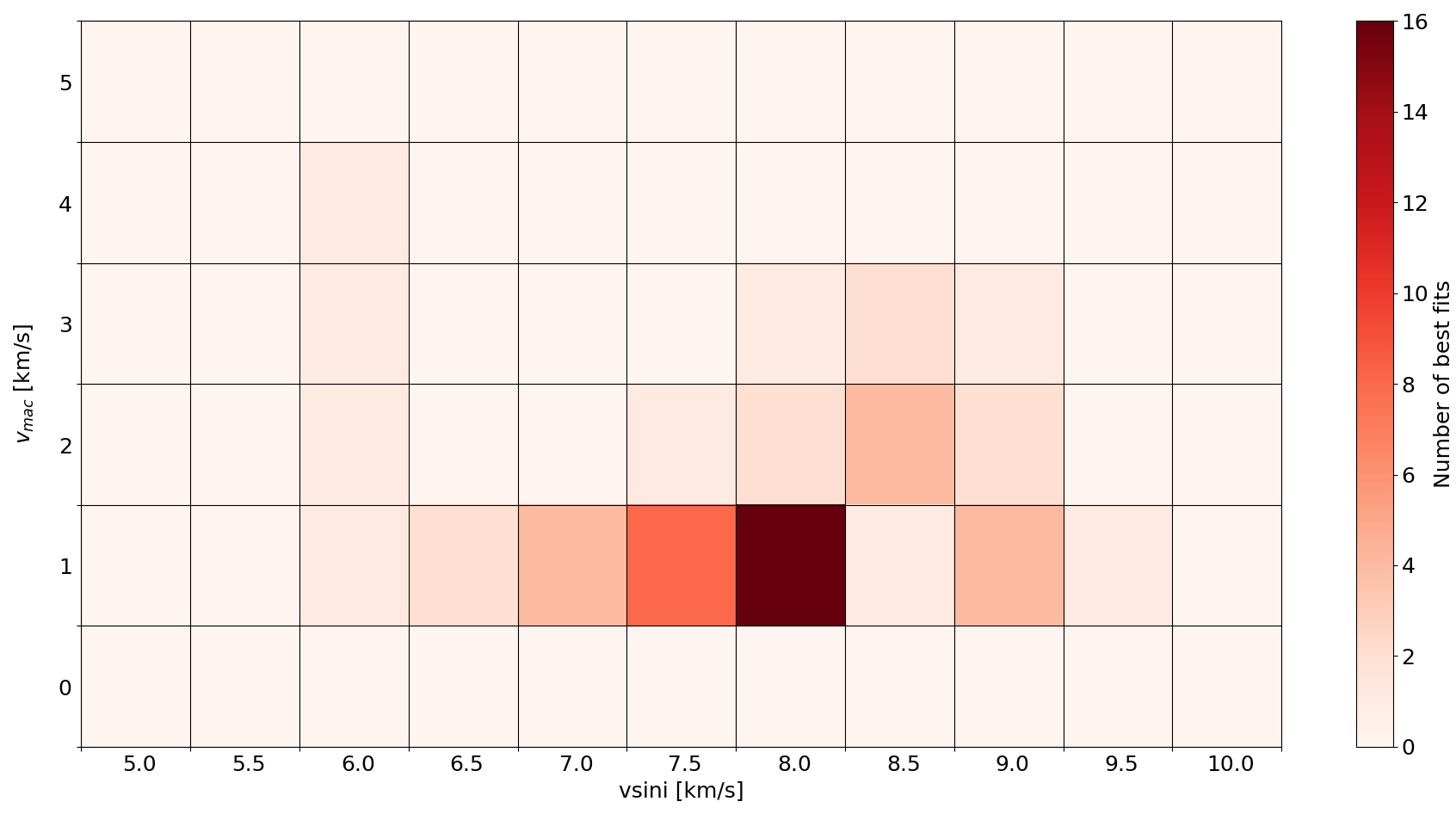}}}
\caption{Distribution of the results for EK Boo with respect to $v\sin i$ and $v_{\rm mac}$.}
\end{figure}
\section{Summary}
An original method is presented for better estimation of the rotational and macroturbulent velocities of cool stars based on a multi-line approach and using spectrum synthesis. A grid of synthetic spectra using MARCS models of atmospheres that take into account a range of $v\sin i$ and $v_{\rm mac}$ is produced. Using the LSD method, a mean spectral line profile is obtained for each synthetic data in the grid and then compared to the mean line profiles obtained from Narval data. Using a minimization rule, the best fitting pair of parameters ($v\sin i$,~$v_{\rm mac}$) is selected for each observation. After inspection, a counting of the individual resulting best fits is done to obtain a statistical final result. Applying the method to the M~giant RZ~Ari, the values $v\sin i=6.0{\pm}0.5$~km/s and $v_{\rm mac}=2.0{\pm}1.0$~km/s are derived. These values are reasonable for the spectral type of RZ~Ari (M6\,III), considering the estimations of Zamanov et al. (2008) as upper limits for M giants, as explained in the Introduction.
\\\\The goal of a future work will be the application of the ZDI method to the same Narval observations using this refined value of $v\sin i$ in order to map the surface magnetic field of RZ~Ari and its temporal evolution. Doing this could help to better understand the origin and evolution of magnetism in this cool evolved star.\\\\
\acknowledgements{We thank the TBL team for providing service observing with Narval. We thank A.~Palacios and S.~Tsvetkova for their ideas, careful reading of the manuscript and constructive comments. S.G., A.L. and R.K.-A. acknowledge partial support by the Bulgarian NSF project DN 18/2, including also observations in semester 2019A. A.L., R.K.-A. and J.M. acknowledge partial support under DRILA 01/3. R.K.-A. acknowledges support for the observational time in 2010 by the Bulgarian NSF project DSAB 01/2. The observations in 2008 and 2011 are under an OPTICON program. The observations in 2013 are under financial support by the OP "Human Resources Development", ESF and the Republic of Bulgaria, project BG051PO001-3.3.06-0047. Since 2015 the Narval observations are under the French "Programme National de Physique Stellaire" (PNPS) of CNRS/INSU co-funded by CEA and CNES. This work has made use of the VALD database, operated at Uppsala University, the Institute of Astronomy RAS in Moscow, and the University of Vienna. We are grateful to the anonymous referee for giving valuable comments and remarks on this paper.}

\end{document}